\documentclass[journal]{IEEEtran}
\ifCLASSINFOpdf
\else
\fi
\usepackage{filecontents}
\usepackage[noadjust]{cite}
\usepackage{multirow}
\usepackage{multicol}
\usepackage{amsfonts}
\usepackage{algorithm}
\usepackage{algorithmic}
\usepackage[table,xcdraw]{xcolor}
\usepackage{makecell}
\definecolor{tabgray}{RGB}{200,200,200}
\usepackage{graphicx}
\usepackage{amsmath}
\usepackage{pgfplotstable}
\usepackage{array}
\usepackage{booktabs}
\usepackage{times}
\usepackage{graphicx}
\usepackage{mathrsfs}
\usepackage{amssymb}
\usepackage{stmaryrd}
\usepackage[noadjust]{cite}
\usepackage{cprotect}

\usepackage[normalem]{ulem} 

\long\def\symbolfootnote[#1]#2{\begingroup%
\def\thefootnote{\fnsymbol{footnote}}\footnote[#1]{#2}\endgroup}

\newcommand{\CASE}[1]{\STATE \textbf{case} #1\textbf{:} \begin{ALC@g}}
\newcommand{\ENDCASE}{\end{ALC@g}}

\newcommand{\DEFAULT}{\STATE \textbf{default:} \begin{ALC@g}}
\newcommand{\ENDDEFAULT}{\end{ALC@g}}
\newcommand{\DEFAULTLINE}[1]{\STATE \textbf{default:} }
\begin{document}
\title{A General Construction and Encoder Implementation of Polar Codes}
\author{
Wei~Song,~\IEEEmembership{Student~Member,~IEEE,} Yifei~Shen,~\IEEEmembership{Student~Member,~IEEE,} Liping~Li,~\IEEEmembership{Member,~IEEE,} Kai~Niu, ~\IEEEmembership{Member,~IEEE,} and~Chuan~Zhang,~\IEEEmembership{Member,~IEEE}
\vspace{-7mm}
\thanks{This work was supported in part by the National Natural Science Foundation
of China through grant 61501002, in part by the Natural Science Project of
Ministry of Education of Anhui through grant KJ2015A102, and
in part by the Talents Recruitment Program of Anhui University.}
\thanks{Wei Song and Liping Li are with the Key Laboratory of Intelligent Computing and Signal Processing,
Ministry of Education,
Anhui University, Hefei, China (liping\_li@ahu.edu.cn).}
\thanks{
Yifei Shen and Chuan Zhang are with the National Mobile Communications Research Laboratory, Southeast University,
Nanjing, China (chzhang@seu.edu.cn)}
\thanks{Kai Niu is with the Key Laboratory of Universal Wireless Communication, Ministry of Education, Beijing Univer-
sity of Posts and Telecommunications, Beijing 100876, Peoples Republic of China
(niukai@bupt.edu.cn)}
}


\maketitle
\begin{abstract}
Similar to existing codes, puncturing and shortening are two general ways to obtain an arbitrary code length and code rate for polar codes.
When some of the coded bits are punctured or shortened, it is equivalent to a situation in
which the underlying channels of the polar
codes are different. Therefore, the quality of bit channels with puncturing or shortening differ from the original qualities,
which can greatly affect the construction of polar codes.
In this paper, a general construction of polar codes is studied in two aspects: 1) the theoretical foundation of the construction; and 2) the hardware implementation of polar codes encoders.
In contrast to the original identical and independent binary-input, memoryless, symmetric (BMS) channels, these underlying BMS channels can be different.
For binary erasure channel (BEC) channels, recursive equations can be employed assuming independent BMS channels.
For all other channel types, the proposed general construction of polar codes is based on the existing Tal-Vardy's procedure.
The symmetric property and the degradation relationship are shown to be preserved under the
general setting, rendering the possibility of a modification of Tal-Vardy's procedure.
Simulation results clearly show improved error performance
with re-ordering using the proposed new procedures.
In terms of hardware, a novel pruned folded encoder architecture is proposed which saves the computation for the beginning frozen bits. Implementation results show the pruned encoder achieve $28\%$ throughput improvement.
\end{abstract}
\section{Introduction}
\IEEEPARstart{P}{olar} codes are proposed by Ar{\i}kan in \cite{arikan_it09} and achieve the capacity of binary-input, memoryless, output-symmetric (BMS) channels with  low encoding and decoding complexity. Given $N$ independent BMS channels $W$, polarization occurs through channel combining and splitting, resulting in perfect bit channels or completely noisy bit channels as $N$ approaches infinity. The portion of the perfect bit channels is exactly the symmetric capacity $I(W)$ of the underlying channel $W$. Polar codes transmit information bits through the perfect bit channels and fix the bits in the completely noisy channels. Accordingly, the bits transmitted through the completely noisy channels are called frozen bits.

The construction of polar codes (selecting the good bit channels from all $N$ bit channels) is presented in \cite{arikan_it09,mori_isit09,mori_icl09,vardy_it13,p.trifonov_c12,dai_ia17,wu_icl14}.
In \cite{arikan_it09}, Ar{\i}kan proposes Monte-Carlo simulations to sort the bit channels with a complexity of  $\mathcal{O}(SN\log N)$ ($S$ represents the iterations of the Monte-Carlo simulations).
In \cite{mori_isit09,mori_icl09}, density evolutions are used in the construction of polar codes.
Since the density evolution includes function convolutions, its precisions are limited by the complexity.
Bit channel approximations are proposed in \cite{vardy_it13} with a complexity of $\mathcal{O}(N\mu^2\log \mu)$ ($\mu$ is a user-defined parameter to control the number of output alphabet at each approximation stage).
In \cite{wu_icl14,dai_ia17,p.trifonov_c12}, the Gaussian approximation (GA) is used to construct polar codes in additive white Gaussian noise (AWGN) channels.

To achieve arbitrary code lengths and code rates, puncturing and shortening of polar codes are reported in \cite{eslami_itp11,niu_icc13,zhang_it14,DBLP:journals/corr/NiuD0LZV16,wang_cl14,vbWCNCW,vmIT}.
In \cite{eslami_itp11}, a channel-independent procedure is proposed for puncturing that involves the minimum stopping
set of each bit. In \cite{eslami_itp11}, the punctured bits are unknown to decoders, and it is therefore called the unknown puncturing type. The  quasi-uniform puncturing (QUP) algorithm is proposed in \cite{niu_icc13}, which simply punctures the reversed bits from 1 to $P$ ($P$ is the number of bits to be punctured). The QUP puncturing is the unknown puncturing. Re-ordering the bit channels after puncturing with the GA method is proposed in \cite{zhang_it14},
and selecting the punctured bits from the frozen positions is also proposed in \cite{zhang_it14}. The puncturing in \cite{zhang_it14} is also the unknown puncturing.
Another type of puncturing, called known puncturing or shortening,
is proposed in \cite{wang_cl14,DBLP:journals/corr/NiuD0LZV16,vbWCNCW,vmIT}.
The reversal quasi-uniform puncturing (RQUP) algorithm proposed in \cite{DBLP:journals/corr/NiuD0LZV16} simply punctures the reversed bits from $N-P+1$ to $N$.
The shortening in \cite{wang_cl14} is based on the column weights of the generator matrices.
A low-complexity construction of shortened and punctured polar codes from a unified view is proposed in \cite{vbWCNCW}.
In \cite{vmIT}, an optimization algorithm to find a shortening pattern and a set of frozen symbols for polar codes is proposed.

Regardless of the puncturing or shortening pattern, a re-ordering
of bit channels is necessary when these operations are performed.
In AWGN channels, GA can be used to re-order the bit channels when some of the coded bits are punctured or shortened.
Puncturing and shortening are equivalent to the case in which
the underlying channel corresponding to the selected coded bit is no
longer the underlying channel $W$. As some of the underlying channels change, the bit channels constructed
from these underlying channels differ from the original channels without puncturing or shortening.
Re-ordering of these new bit channels is necessary to avoid  deterioration of performance.
However, the GA method is only applicable to AWGN channels. New procedures are important when studying puncturing or shortening of polar codes. This is the motivation of the work in this paper.

To study the construction of polar codes in which some of the coded bits are punctured or shortened,
we first generalize this problem by considering the underlying channels to be independent BMS channels (not necessarily identical ones).
For BEC channels, recursive equations are proposed in \cite{korada_thesis} to calculate the Bhattacharyya parameter for each bit channel. The construction complexity is the same as the original complexity in \cite{arikan_it09}.
For other channel types, the general construction in this paper is based on Tal-Vardy's procedures in \cite{vardy_it13}.
The symmetric property of polar codes, which is first stated in \cite{arikan_it09}, is proven to  hold
in the new setting in which the underlying channels can differ. The degradation relationship (the foundation of
Tal-Vardy's procedure) is also proven to hold.
Based on the theoretical analysis, a modification to the Tal-Vardy algorithm \cite{vardy_it13} that is applicable to
any BMS channel is proposed to re-order the bit channels when some of the underlying channels are independent BMS
channels (which, again, could be different channels). For continuous output channels such as AWGN channels, conversion to BMS channels can be performed first, and then the modified Tal-Vardy algorithm can be applied analogous to the Tal-Vardy algorithm itself.
The general construction can therefore be applied to re-order the bit channels with puncturing or shortening.
Depending on the puncture
type, the punctured channel must be equivalently modeled. Then, the recursion in BEC channels or the modified Tal-Vardy's
procedure for all other channels can be applied to re-order the bit channels.
Simulation results show that the re-ordering greatly improves the error performance of polar codes.

Utilizing the property that the beginning of the source bits are usually frozen bits ($0$s in other words), the encoding throughput can be improved. With the increase of the code length, the area of the encoder increases exponentially. Folding \cite{parhi2007vlsi} is a technique to reduce the area by multiplexing the modules. By exploiting the same property between polar encoding and the fast Fourier transformation (FFT), \cite{yoo2015partially} first applies the folding technique for the polar encoding based on \cite{ayinala2012pipelined}. Folded systematic polar encoder is implemented in \cite{sarkis2015flexible}. Moreover, \cite{zhong2018auto} designs an auto-generation folded polar encoder, which could preprint the hardware code directly given the length and the level of parallelism. Combining the property of the puncturing mode, current folded encoder could be pruned further. In this paper, a pruned folded polar encoder is proposed. It avoids the beginning calculation of the frozen `$0$' bits. Therefore, the latency could be reduced significantly. Implementation results also proves the feasibility of the pruned encoder, which provides $28\%$ throughput improvement.

The remainder of this paper is organized as follows. In Section \ref{sec.pre1}, we briefly introduce the basics of polar codes.
The general construction based on Tal-Vardy's procedure is presented in Section \ref{sec.vardy}.
The numerical results for applying the BEC construction and the general construction in Section \ref{sec.vardy} are
provided in Section \ref{sec_numerical}. Section \ref{sec:vlsi} proposes a pruned folded polar encoder architecture and the results are compared with the state-of-the-art. The paper ends with concluding remarks.

\section{Background  on Polar Codes}\label{sec.pre1}
\subsection{Polarization Process}\label{sub.2.1}
For a given BMS channel $W$: $\mathcal{X}$ $\longrightarrow$ $\mathcal{Y}$, its input alphabet, output alphabet, and transition probability are $\mathcal{X}=\{0,1\}$, $\mathcal{Y}$, and $W(y|x)$, respectively, where $x\in\mathcal{X}$ and $y\in\mathcal{Y}$. Two parameters represent the quality of a BMS channel $W$: the symmetric capacity and the Bhattacharyya parameter. The symmetric capacity can be expressed as
\begin{equation}\label{eq_iw}
I(W)=\sum\limits_{y\in \mathcal{Y}}\sum\limits_{x\in \mathcal{X}} \frac{1}{2}W(y|x)\log\frac{W(y|x)}{\frac{1}{2}W(y|0)+\frac{1}{2}W(y|1)}.
\end{equation}
The  Bhattacharyya parameter is
\begin{equation}
Z(W)=\sum\limits_{y\in \mathcal{Y}} \sqrt{W(y|0)W(y|1)}\,.
\end{equation}

The term $G_N$ is used to represent the generator matrix: $G_N = B_NF^{\otimes n}$, where $N = 2^n$ is the  code length  ($n\geqslant 1$), $B_N$ is the permutation matrix used for the bit-reversal operation,
$F\triangleq[\begin{smallmatrix}1&0\\1&1 \end{smallmatrix}]$,
and $F^{\otimes n}$ denotes the $n$th Kronecker product of $F$.
The channel polarization is divided into two phases: channel combing and channel splitting. The channel combing refers to the combination of $N$ copies of a given BMS $W$  to produce a vector channel $W_N$,  defined as
\begin{align}
W_N(y_1^N|u_1^N)=W^N(y_1^N|u_1^NG_N).
\end{align}
The channel splitting splits $W_N$ back into a set of $N$ binary-input  channels $W_N^{(i)}$, defined as
\begin{align}\label{eq.WN}
W_N^{(i)}(y_1^N,u_1^{i-1}|u_i)=\sum\limits_{u_{i+1}^N\in \mathcal{X}^{N-i}}\frac{1}{2^{N-1}}W_N(y_1^N|u_1^N).
\end{align}
The channel $W_N^{(i)}$ is called bit channel $i$, {{which indicates}} that it is the channel that bit $i$ experiences from the channel combining and splitting stages. {{Bit channel $i$ can be viewed as a BMS channel: $\mathcal{X} \rightarrow (\mathcal{X}_1^{i-1},\mathcal{Y}_1^N)$.}}

\begin{figure}
{\par\centering
\resizebox*{3.0in}{!}{\includegraphics{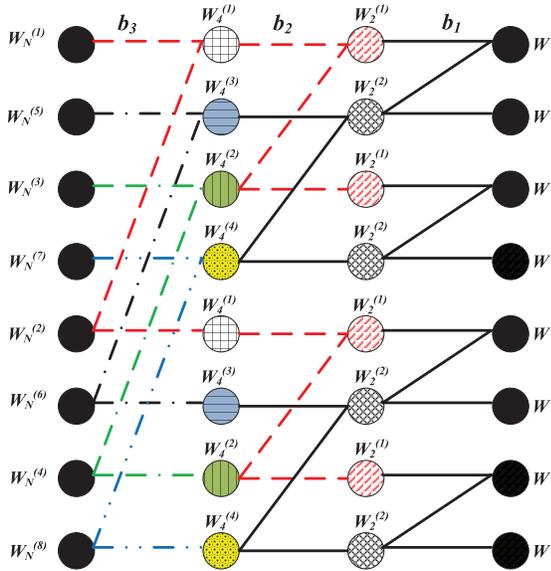}} \par}
\caption{Full expansion of the tree structure of polar codes construction for $N=8$. At level 1, all Z-shapes
have the same input channel. At level 2, there are 2 groups of Z-shapes, each with their own input channels: the
group containing the dashed-line Z-shapes and the group containing the solid-line Z-shapes. At level 3, there are 4 groups of Z-shapes, each with different input channels fed from the output channels of level 2.}
\label{fig.Zshapes}
\end{figure}

Polar codes can also be constructed recursively in a tree structure \cite{arikan_it09}. The tree structure is expanded fully
in Fig.~\ref{fig.Zshapes} for $N=8$. There are eight independent and identical BMS channels $W$ at the right-hand side.
In Fig.~\ref{fig.Zshapes}, from right to left, there are three levels: level one, level two, and level three, each
containing $N/2$ Z-shapes. A Z-shape is the basic one-step transformation with the transition probability defined in (\ref{eq.WN}) with $N=2$. This one-step transformation converts two input channels to two output channels: the upper left
channel and the lower left channel.
For bit channel $i$ ($1\le i \le N$),  the binary expansion of $i-1$  is denoted as
$\langle i \rangle = (b_1, b_2, ..., b_n)$ ($b_1$ being the MSB). The bit $b_k$ at level $k$ ($1 \le k \le n$)
determines whether bit channel $i$ takes
the upper left channel or the lower left channel: If $b_k = 0$, bit channel $i$ takes the upper left channel; otherwise, it takes the
lower left channel. At level $k$, there are $2^{n-k}$ Z-shapes with the same input channels.
For example, in Fig.~\ref{fig.Zshapes}, at
level $1$, all Z-shapes have the same input channels $W$. At level $2$, there are two Z-shapes with the same input channels:
the two dashed-line Z-shapes  have  input channels $W_2^{(1)}$, and the two solid-line Z-shapes have input channels $W_2^{(2)}$.
The Z-shapes are grouped with the same input channels as one group in each level. Then, at level 1, there is one group
(containing four Z-shapes)
sharing the same input channels. By contrast,  at level 2, there are two groups (each containing two Z-shapes) that share the
same input channels (the dashed-line group and the solid line group). At level 3, all four
Z-shapes have different input channels.
To construct polar codes, the one-step
transformation of the Z-shapes in the same group only needs to be calculated once \cite{arikan_it09,vardy_it13}.

\subsection{Motivation for the General Construction}\label{sub.2.2}
 The original code  length $N$  of polar codes is limited to the power of two, i.e.,  $N = 2 ^ n$.
 To obtain any code length, puncturing or shortening is typically performed.
  For the puncturing mode, some coded bits are punctured in the encoder and the decoder has no a priori information about these bits. For the shortening mode, the values of the shortened coded bits are known by both the encoder and decoder.

 The code lengths of both the punctured codes and the shortened codes are denoted by $M$. Let $P$ denote the number of punctured (or shortened) bits, with $P=N-M$. The code rate of the punctured or shortened codes is $R$.
 For the punctured mode, the decoder does not have a priori information on the punctured bits.
 {{
 A BMS punctured channel of this type can be modeled as a BMS channel $H$ with $H(y|0) = H(y|1)$
 since for the received symbol $y$,
 the likelihood of $0$ or $1$ being transmitted is equal. The following lemma can be easily checked.
 \newtheorem{lemma}{Lemma}
 \begin{lemma}\label{lemma_puncture_channel}
 For a punctured channel $H: \mathcal{X} \rightarrow \mathcal{Y}$ with
 $H(y|0) = H(y|1)=1/2$ ($y \in \mathcal{Y}$ and $|\mathcal{Y}| = 1$), the symmetric capacity of $H$ is $I(H) = 0$.
 \end{lemma}
 The proof of this lemma can be found in the Appendix.

 For the shortened mode, the shortened bits are  known to the decoder that can be modeled from the
 following lemma.

 \begin{lemma}\label{lemma_shortened_channel}
 A shortened channel with shortened bits known to the receiver can be modeled as a binary
 symmetric channel (BSC) $H'$ with a cross-over transition probability zero: $H': \mathcal{X} \rightarrow \mathcal{Y}$ with $H'(0|0) = H'(1|1) = 1$ and $H'(0|1) = H'(1|0) = 0$. The capacity of a shortened channel $H'$ is therefore $I(H') = 1$.
 \end{lemma}
 The focus of this lemma is the model of the shortened channel $H'$. The capacity of $H'$ is a well-known result \cite{cover_it}.
 }}



\begin{figure}
{\par\centering
\resizebox*{3.0in}{!}{\includegraphics{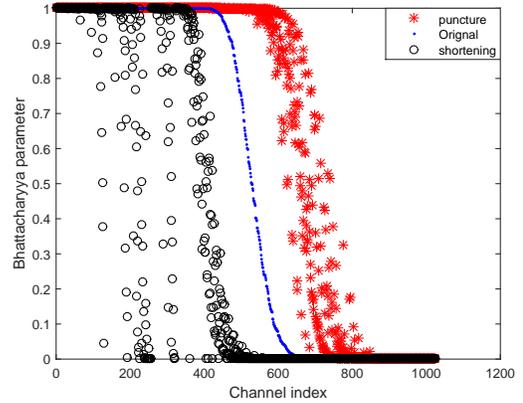}} \par}
\caption{Comparison of the Bhattacharyya parameters of the bit channels between the original polar code, the punctured polar code and the shortened polar code with $N=1024$ and $M=700$ in the BEC channel with an erasure probability of 0.5.}
\label{fig.BEC-Paste parameter change}
\end{figure}

 Once some of the coded bits are punctured or shortened, the underlying channels are no longer the same channels as originally proposed in \cite{arikan_it09}. The bit channels constructed from the channel combining and splitting stages therefore have different qualities and must be re-ordered. Fig.~\ref{fig.BEC-Paste parameter change} shows an example of the Bhattacharyya parameters of bit channels constructed from a underlying BEC channel with an erasure probability 0.5. In Fig.~\ref{fig.BEC-Paste parameter change}, the original polar code block length is $N=1024$. The blue dots are the Bhattacharyya parameters of the original bit channels. The red asterisks are the Bhattacharyya parameters of the bit channels with $P=324$ punctured coded bits,  and these punctured bits are unknown to the receiver. Equivalently, among the original $N=1024$ independent BEC channels, $P=324$ channels are now completely noisy channels ($I(W)=0$). The Bhattacharyya parameters of the bit channels in this case are worse than the original bit channels, as indicated by the red asterisks in Fig.~\ref{fig.BEC-Paste parameter change}. The black circles are the Bhattacharyya parameters of the bit channels with $P=324$ shortened coded bits,  and these shortened bits are known to the receiver. Equivalently, among the original $N=1024$ independent BEC channels, $P=324$ channels are now completely perfect channels ($I(W)=1$).
 Therefore, the good bit channels should be re-selected from the new set of bit channels with different Bhattacharyya parameters due to puncturing or shortening.


\subsection{Construction of BEC Channels}
\begin{figure}[!t]
{\par\centering
\resizebox*{2.4in}{!}{\includegraphics{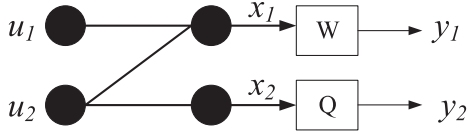}} \par}
\caption{General one-step transformation  of polar codes where the underlying channels are independent BMS channels.}
\label{fig.basic structure}
\end{figure}

First, we consider the one-step transformation and generalize the original transformation from two identical independent underlying channels $W$ to two independent underlying channels. The two independent channels can be different channels as indicated in Fig.~\ref{fig.basic structure}, where $W$ and $Q$ are two BMS channels. With puncturing or shortening, $Q$ can be a completely noisy channel with $I(Q)=0$ (puncturing) or a perfect channel $I(Q)=1$ (shortening). With the generalization in  Fig.~\ref{fig.basic structure}, the synthesized channel $W_2$ can be expressed as follows:
\begin{align}
W_2(y_1,y_2|u_1, u_2)=W(y_1|u_1\oplus u_2)Q(y_2|u_2).\label{eq.W2}
\end{align}
The splitting channels $W_2^{(1)}$ and $W_2^{(2)}$ can be expressed as follows:
\begin{align}
W_2^{(1)}(y_1^2|u_1)=\sum\limits_{u_2}\frac{1}{2}W(y_1|u_1\oplus u_2)Q(y_2|u_2),\label{eq.W21}\\
W_2^{(2)}(y_1^2,u_1|u_2)=\frac{1}{2}W(y_1|u_1\oplus u_2)Q(y_2|u_2).\label{eq.W22}
\end{align}
Borrowing the notations from \cite{korada_thesis}, the above one-step transformation can be
written as follows:
\begin{align}
W_2^{(1)}=W \boxast Q, \label{eq_W21_simple}\\
W_2^{(2)}= W \varoast Q. \label{eq_W22_simple}
\end{align}

For BEC channels,  the one-step transformation from $W$ and $Q$ to $W_2^{(1)}, W_2^{(2)}$ has the following Bhattacharyya parameters \cite{korada_thesis}:
\begin{align}
Z(W_2^{(1)})=Z(W)+Z(Q)-Z(W)Z(Q),\label{eq.BEC_1}\\
Z(W_2^{(2)})=Z(W)Z(Q).\label{eq.BEC_2}
\end{align}
The construction of polar codes in BEC channels can be performed by recursively employing these two equations
where the underlying channels are independent BMS channels.

\section{General Construction Based on \\ the Tal-Vardy Procedure}\label{sec.vardy}
Tal-Vardy's construction of polar codes \cite{vardy_it13} is based on the fact that polar codes can be
constructed in $n$ levels for a block length $N=2^n$, as shown in Fig.~\ref{fig.Zshapes}.
In each level, the one-step transformation defined in (\ref{eq.W21}) or (\ref{eq.W22}) is performed.
Note that in the original
one-step transformation in \cite{arikan_it09}, the underlying channels are identical: $Q = W$ (also shown
in Fig.~\ref{fig.Zshapes}).
From level $k$ to level $k+1$ ($1 \le k \le n-1$), the size of
the output alphabet at least squares. The output channel in each level is still a BMS channel.
The idea in \cite{vardy_it13} is to approximate
the output BMS channel in each level by a new BMS channel with a controlled output alphabet size.
This size is denoted as $\mu$, which indicates that the output alphabet has at most $\mu$ symbols in each level.
With this controlled size, each approximated bit channel can be evaluated in terms of the error probability,

As shown in \cite{vardy_it13,korada_thesis},  the degradation relation
is preserved by the one-step channel transformation operation. In addition, as shown by Proposition 6
of \cite{vardy_it13},  the output of
the approximate procedure remains a BMS channel: taking an input BMS channel, the
output from the approximate process is still a BMS channel.
Therefore, the key to applying Tal-Vardy's approximate function to the general construction
is that 1) the output channels from (\ref{eq.W21}) and (\ref{eq.W22}) are still BMS channels; 2) the degradation relation is preserved from (\ref{eq.W21}) and (\ref{eq.W22}).

In the following,
we first prove that the output of the generalized one-step transformation remains a BMS channel. Then, the degradation relation from (\ref{eq.W21}) and (\ref{eq.W22}) is shown to be preserved.
The modification to Tal-Vardy's algorithm follows.

\subsection{Symmetric Property}
Some notations are needed first.
For a BMS channel $W$, it has a permutation $\pi_1$ on $\mathcal{Y}$ with $\pi_1^{-1}$=$\pi_1$ and $W(y|1)=W(\pi_1(y)|0)$ for all $y\in \mathcal{Y}$. Let $\pi_0$ be the identity permutation on $\mathcal{Y}$. As in \cite{arikan_it09}, a simpler expression $x \cdot y$ is used to replace $\pi_x(y)$, for $x\in \mathcal{X}$, and $y\in \mathcal{Y}$.

Obviously, the equation $W(y|x\oplus a)=W(a\cdot y|x)$ is established for $a\in \{0,1\}$, $x\in \mathcal{X}$, and $y\in \mathcal{Y}$. It is also shown in \cite{arikan_it09} that $W(y|x\oplus a)=W(a\cdot y|x)$.
For $x_1^N\in \mathcal{X}^N$ and $y_1^N \in\mathcal{Y}^N$, let

  \begin{equation}
    x_1^N\cdot y_1^N\triangleq (x_1\cdot y_1,...,x_N\cdot y_N).
  \end{equation}
This is an element-wise permutation.
\newtheorem{proposition}{Proposition}
\begin{proposition}\label{proposition3}
For two independent BMS channels $W$ and $Q$, $W$$^2$=$W\ast Q$ is also symmetric in the sense that
\begin{equation}
W^2(y_1^2|x_1^2\oplus a_1^2)=W^2(x_1^2\cdot y_1^2|a_1^2),\label{eq.14}
\end{equation}
where $a_1,a_2 \in \{0,1\}$.
\end{proposition}
\begin{IEEEproof}
Observe that $W^2(y_1^2|x_1^2\oplus{a_1^2})=W(y_1|x_1\oplus a_1)Q(y_2|x_2\oplus a_2)$.
From the symmetric property of $W$ and $Q$, we also have
$$W(y_1|x_1\oplus a_1)=W(x_1\cdot y_1|a_1),$$
$$Q(y_2|x_2\oplus a_2)=Q(x_2\cdot y_2|a_2).$$
Therefore, $W^2(y_1^2|x_1^2\oplus a_1^2)=W(x_1\cdot y_1|a_1)Q(x_2\cdot y_2|a_2)$,
which is exactly $W^2(x_1^2\cdot y_1^2|a_1^2)$.
\end{IEEEproof}
\begin{proposition}\label{proposition4}
For BMS channels $W$ and $Q$, the channels $W_2$ defined in (\ref{eq.W2}), $W_2^{(1)}$ defined in (\ref{eq.W21}), and $W_2^{(2)}$ defined in (\ref{eq.W22}), are also symmetric in the sense that
\begin{align}
W_2(y_1^{2}|u_1^{2})=W_2(a_1^{2}G_2\cdot y_1^{2}|u_1^{2}\oplus a_1^{2}),\label{eq.15}\\
W_2^{(1)}(y_1^{2}|u_1)=W_2^{(1)}(a_1^2G_2\cdot y_1^{2}|u_1\oplus a_1),\label{eq.16}\\
W_2^{(2)}(y_1^{2},u_1|u_2)=W_2^{(2)}(a_1^2G_2\cdot y_1^2,u_1\oplus a_1|u_2\oplus a_2),\label{eq.17}
\end{align}
where $a_1,a_2 \in \{0,1\}$.
\end{proposition}
\begin{IEEEproof}
Let $x_1^2=u_1^2G_2$ and observe that
\begin{align*}
W_2(y_1^2|u_1^2)&=W(y_1|x_1) Q(y_2|x_2)\\
&=W(x_1\cdot y_1|0) Q(x_2\cdot y_2|0)\\
&=W_2(x_1^2\cdot y_1^2|0_1^2).
\end{align*}
Let $b_1^2=a_1^2G_2$, and observe that
\begin{align*}
&W_2(a_1^2G_2\cdot y_1^2|u_1^2\oplus a_1^2)=W^2(b_1^2\cdot y_1^2|(u_1^2\oplus a_1^2)G_2)\\
&=W^2(b_1^2\cdot y_1^2|(u_1^2G_2)\oplus (a_1^2G_2))\\
&=W^2(b_1^2\cdot y_1^2|x_1^2\oplus b_1^2)\\
&=W(b_1\cdot y_1|x_1\oplus b_1) Q(b_2\cdot y_2|x_2\oplus b_2)\\
&=W_2(x_1^2\cdot y_1^2|0_1^2).
\end{align*}
This proves the first result in (\ref{eq.15}).
Next, we prove the second claim in (\ref{eq.16}).
With $x_1=u_1\oplus u_2$ and $x_2=u_2$, the bit channel $W_N^{(1)}$ can be written as follows:
\begin{align} \nonumber
W_2^{(1)}(y_1^2|u_1)&=\sum\limits_{u_2}\frac{1}{2}W_2(y_1^2|u_1^2)\\
&=\sum\limits_{u_2}\frac{1}{2}W(y_1|x_1) Q(y_2|x_2)\\ \nonumber
&=\sum\limits_{u_2}\frac{1}{2}W(x_1\cdot y_1|0) Q(x_2\cdot y_2|0)\\ \nonumber
&=\sum\limits_{u_2}\frac{1}{2}W_2(x_1^2\cdot y_1^2|0_1^2).
\end{align}
Then, let $b_1^2=a_1^2G_2$, and observe that
\begin{align*}
&W_2^{(1)}(a_1^2G_2\cdot y_1^2|u_1\oplus a_1)\\
&\quad\quad=\sum\limits_{u_2\oplus a_2}\frac{1}{2}W_2(b_1^2\cdot y_1^2|u_1^2\oplus a_1^2)\\
&\quad\quad=\sum\limits_{u_2}\frac{1}{2}W((x_1\oplus b_1)\cdot (b_1\cdot y_1)|0)\\
&\quad\quad\quad\quad \times Q((x_2\oplus b_2)\cdot(b_2\cdot y_2)|0)\\
&\quad\quad=\sum\limits_{u_2}\frac{1}{2}W_2(x_1^2\cdot b_1^2\cdot b_1^2\cdot y_1^2|0_1^2)\\
&\quad\quad=\sum\limits_{u_2}\frac{1}{2}W_2(x_1^2\cdot y_1^2|0_1^2).
\end{align*}
Thus, the second claim in (\ref{eq.16}) is established.
Finally, we prove the final claim. Observe that
\begin{equation}
\begin{split}
W_2^{(2)}(y_1^2,u_1|u_2)&=\frac{1}{2}W(y_1|x_1) Q(y_2|x_2)\nonumber\\
&=\frac{1}{2}W(x_1\cdot y_1|0) Q(x_2\cdot y_2|0)\nonumber\\
&=\frac{1}{2}W_2(x_1^2\cdot y_1^2|0_1^2).
\end{split}
\end{equation}
Then, observe that
\begin{align*}
&W_2^{(2)}(a_1^2G_2\cdot y_1^2,u_1\oplus a_1|u_2\oplus a_2)\\
&\quad\quad=\frac{1}{2}W(b_1\cdot y_1|x_1\oplus b_1) Q(b_2\cdot y_2|x_2\oplus b_2)\\
&\quad\quad=\frac{1}{2}W((x_1\oplus b_1)\!\cdot\!(b_1\cdot y_1)|0)\times Q((x_2\oplus b_2)\!\cdot\!(b_2\!\cdot\! y_2)|0)\\
&\quad\quad=\frac{1}{2}W(x_1\cdot b_1\cdot b_1\cdot y_1|0) Q(x_2\cdot b_2\cdot b_2\cdot y_2|0)\\
&\quad\quad=\frac{1}{2}W(x_1\cdot y_1|0) Q(x_2\cdot y_2|0)\\
&\quad\quad=\frac{1}{2}W_2(x_1^2\cdot y_1^2|0_1^2).
\end{align*}
Thus, the third claim in (\ref{eq.17}) is also established.
\end{IEEEproof}
From equations (\ref{eq.14}), (\ref{eq.15}), (\ref{eq.16}) and (\ref{eq.17}), it can be seen  that  $W^2$, $W_2$,
$W_2^{(1)}$ and $W_2^{(2)}$ are still symmetric channels with two independent underlying BMS channels $W$ and $Q$.

\subsection{Degradation Relation}
First, the notation $\preccurlyeq$ denotes the degraded relationship as in \cite{korada_thesis}.
Therefore, degradation of $Q$ with respect to $W$ can be denoted as $Q\preccurlyeq W$.

\begin{lemma}\label{lemma1}
Let $H$ and $Q$ be two underlying BMS channels.
The output channels $H_2^{(1)}$ and $H_2^{(2)}$ constructed from these channels are defined in (\ref{eq.W21}) and (\ref{eq.W22}).
Let $W_2^{(1)}$ and $W_2^{(2)}$ be constructed from two independent underlying BMS channels $W$ and $Q$. If $H\preccurlyeq W$, then
\begin{equation}
H_2^{(1)}\preccurlyeq W_2^{(1)}\quad\quad  {\text{and}}  \quad\quad H_2^{(2)}\preccurlyeq W_2^{(2)}.
\end{equation}
\end{lemma}
\begin{IEEEproof}
Let $P_1$: $\mathcal{Y}\rightarrow\mathcal{Z}$ be an intermediate channel that degrades $W$ to $\mathcal{Z}$.
That is, for all $z\in \mathcal{Z}$ and $x\in \mathcal{X}$, we have
\begin{align}\label{eq.degrated}
H(z|x)=\sum\limits_{y\in \mathcal{Y}}W(y|x)P_1(z|y).
\end{align}
We first prove the first part of this lemma: $H_2^{(1)}\preccurlyeq W_2^{(1)}$.
According to (\ref{eq.W21}),  $H_2^{(1)}$ can be written as
\begin{align}\label{eq_H21}
H_2^{(1)}(z_1,z_2|u_1)=\frac{1}{2}\sum\limits_{u_2\in \mathcal{X}}H(z_1|u_1\oplus u_2)Q(z_2|u_2).
\end{align}
Apply (\ref{eq.degrated}) to (\ref{eq_H21}):
\begin{align}\label{eq.d1}
&H_2^{(1)}(z_1,z_2|u_1)=\nonumber\\
&\frac{1}{2}\sum\limits_{u_2\in \mathcal{X}}\sum\limits_{y_1\in \mathcal{Y}}W(y_1|u_1\oplus u_2)P_1(z_1|y_1)Q(z_2|u_2).
\end{align}
As shown in \cite{vardy_it13}, a channel can be considered to be both degraded and upgraded with respect to itself.
Let $P_2$ be the channel to degrade $Q$ to itself:
\begin{align}\label{eq.degraded1}
Q(z|x)=\sum\limits_{y\in \mathcal{Y}}Q(y|x)P_2(z|y).
\end{align}
Next, apply (\ref{eq.degraded1}) to (\ref{eq.d1}):
\begin{align}
&H_2^{(1)}(z_1,z_2|u_1)=\frac{1}{2}\sum\limits_{u_2\in \mathcal{X}}\sum\limits_{y_1^2\in \mathcal{Y}^2}W(y_1|u_1\oplus u_2)Q(y_2|u_2)\nonumber\\
&\quad\quad\quad\quad\quad\quad\quad ~~~~~~~~~~~~~~~~~~\times P_1(z_1|y_1)P_2(z_2|y_2)\nonumber\\
&\quad\quad\quad \quad\quad\quad =\sum\limits_{y_1^2\in \mathcal{Y}^2}W_2^{(1)}(y_1,y_2|u_1)P_1(z_1|y_1)P_2(z_2|y_2).\label{eq.5}
\end{align}
Define another channel $P^*$ as $\mathcal{Y}^2\rightarrow\mathcal{Z}^2$ for all $y_1^2\in \mathcal{Y}^2$ and $z_1^2 \in \mathcal{Z}^2$:
\begin{align}\label{eq.imme}
P^*(z_1^2|y_1^2)=P_1(z_1|y_1)P_2(z_2|y_2).
\end{align}
Applying (\ref{eq.imme}) to (\ref{eq.5}), we have
\begin{align*}
H_2^{(1)}(z_1,z_2|u_1)=\sum\limits_{y_1^2\in \mathcal{Y}^2}W_2^{(1)}(y_1,y_2|u_1)P^*(z_1^2|y_1^2),
\end{align*}
which shows that $H_2^{(1)}$ is degraded with respect to $W_2^{(1)}$: $H_2^{(1)} \preccurlyeq W_2^{(1)}$. This proves the first
part of this lemma.
In the same fashion, the second part of the lemma can be proven: $H_2^{(2)}\preccurlyeq W_2^{(2)}$.
\end{IEEEproof}
\begin{lemma}\label{lemma2}
Let $H$ and $T$ be two underlying BMS channels.
The output channels $H_2^{(1)}$ and $H_2^{(2)}$ are defined in (\ref{eq.W21}) and (\ref{eq.W22}).
Let $W_2^{(1)}$ and $W_2^{(2)}$ be constructed from two independent underlying BMS channels $W$ and $Q$.
If $H\preccurlyeq W$ and $T\preccurlyeq Q$, then
\begin{equation}
H_2^{(1)}\preccurlyeq W_2^{(1)}\quad\quad  {\text{and}}  \quad\quad H_2^{(2)}\preccurlyeq W_2^{(2)}.
\end{equation}
\end{lemma}
The proof of this lemma is immediately available following the proof of Lemma \ref{lemma1}.
{{Please note that the proof of Lemma \ref{lemma1} and Lemma \ref{lemma2} is provided in \cite{alsan_thesis}}, where
the increasing convex ordering property is invoked. In this part, we provide another way
to prove the degradation preservation of the general one-step polar transformation.}

\begin{proposition}
Suppose there is a degrading (upgrading) algorithm that approximates the BMS channel of the
one-step transformation defined in (\ref{eq.W21}) or (\ref{eq.W22}) with another BMS channel.  Apply
the approximate algorithm to each of the $n$ one-step transformations. For the $i$th ($1\le i \le N$) bit channel,
denote the final approximate bit channel as ${W'}_N^{(i)}$. Then, ${W'}_N^{(i)}$ is a BMS channel that is
 degraded (upgraded) with respect to $W_N^{(i)}$.
\end{proposition}

\begin{IEEEproof}
From Proposition \ref{proposition3} and Proposition \ref{proposition4}, it is shown that the output of the one-step transformation is still a BMS channel. From Lemma \ref{lemma1} and Lemma \ref{lemma2}, it is shown
that the degradation relation is preserved with the transformation defined in (\ref{eq.W21}) and (\ref{eq.W22}).
Then with induction on each one-step transformation, the final bit channel applying any degrading (upgrading) algorithm
is a degraded (upgraded) version of the original bit channel.
\end{IEEEproof}

In the following subsection, the approximate procedure is chosen to be  Tal-Vardy's in \cite{vardy_it13}.

\subsection{Modified Tal-Vardy Algorithm} \label{sec_modified_construction}
The Tal-Vardy algorithm is used to construct polar codes in \cite{vardy_it13}.
The algorithm can obtain an approximating bit-channel with a specific size $\mu$ using the degrading merge function or the upgrading merge function.
The underlying channels are assumed to be independent and identical BMS channels.
In this part, we propose a modification to Tal-Vardy's approximate procedure that can take independent BMS underlying channels.

\begin{figure}
{\par\centering
\resizebox*{3.0in}{!}{\includegraphics{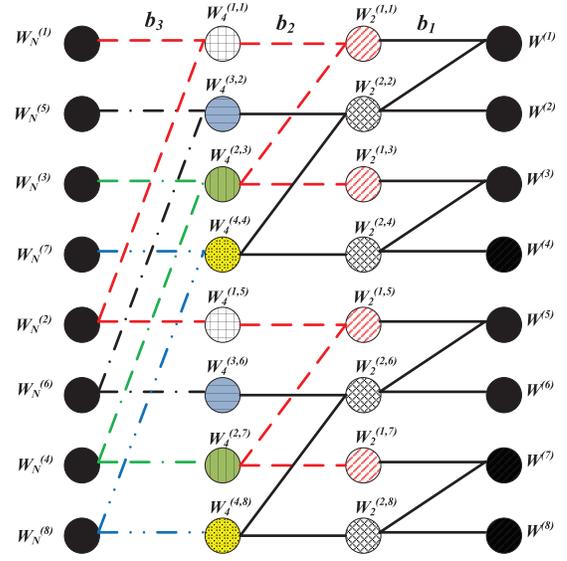}} \par}
\caption{The general construction of polar codes for $N=8$. The initial $N=8$ bit channels are independent BMS
channels $W^{(1)}$, $W^{(2)},\ldots,W^{(8)}$. The output channels of level 1 are labeled  $W_2^{(1,j)}$ and
$W_2^{(2,j)}$ ($1 \le j \le 8$) to indicate that they could be different. The same labeling is applied to the output channels of level 2. }
\label{fig.Zshapes_general}
\end{figure}

Fig.~\ref{fig.Zshapes_general} shows an example of the general construction, indicating the
key difference of the construction problem with the original construction.
The labeling of the intermediate channels in Fig.~\ref{fig.Zshapes_general}
is different from
the labeling in Fig.~\ref{fig.Zshapes}. For example, the four channels $W_2^{(1)}$ (these
are output channels of level one) in Fig.~\ref{fig.Zshapes} are
identical BMS channels, whereas in Fig.~\ref{fig.Zshapes_general}, these channels
could be independent but different BMS channels. The superscript $W_2^{(1)}$ is changed to $W_2^{(1,j)}$ to differentiate
these channels, where $j$ ($1 \le j \le N$) is the position of the channel counting from the top to the bottom in that level.
Originally, there are four channels $W_2^{(1)}$ located at positions 1, 3, 5, and 7.
These  are now possibly different channels: $W_2^{(1,1)}$,
$W_2^{(1,3)}$, $W_2^{(1,5)}$, and $W_2^{(1,7)}$ in the general construction.
The labelling of the output channels at level 2 follows the same fashion. Originally,
two Z-shapes composed of  the four channels $W_2^{(1)}$ belong to the same group, thus requiring only one calculation
of the one-step transformation. In the general construction, since $W_2^{(1,1)}$,
$W_2^{(1,3)}$, $W_2^{(1,5)}$, and $W_2^{(1,7)}$ could be different, the two Z-shapes composed of them need to be
evaluated.

We use \verb"approximateFun"$(W,\mu)$ to represent the degrading or the upgrading procedure in \cite{vardy_it13}.
The vector $W$ contains $N$ sections: $W=(W^{(1)}, W^{(2)}, \ldots, W^{(N)})$, with $W^{(i)}$ representing the transition probability of the $i$th underlying channel.
As in \cite{vardy_it13}, suppose $W^{(i)}$ is sorted according to the ascending order of the likelihood ratios.
The modified Tal-Vardy procedure is presented in Algorithm \ref{al.ag2}. Algorithm \ref{algorithm_transition_prob}
locates the transition probabilities of the two underlying channels for a given Z-shape at a given level.

\begin{algorithm}
\caption{Modified Tal-Vardy Algorithm}
\label{al.ag2}
\begin{algorithmic}[1]
\REQUIRE $n$: block length $N=2^n$ \\
~~~~~$\mu$: the size of the output channel alphabet \\
~~~~$W$: transition probability of the underlying channel $W=(W^{(1)}, W^{(2)},\ldots, W^{(N)})$;
\ENSURE $P_e$;$//$ The vector containing the error probability of bit channels from 1 to $N$.
\FOR{$i=1$ to $n$}
\FOR{$j=1$ to $\frac{N}{2}$}
\STATE $(W_u, W_b, k_1, k_2)=$\verb"tran"$(W,i,j)$; 
\STATE $//$ Obtain the transition probability defined in (\ref{eq.W21})
\STATE $W_0\leftarrow$ \verb"calcTran_typeZero"$(W_u, W_b);$
\STATE $//$ Merge  $W_{0}$ with a fixed output alphabet size $\mu$
\STATE $W_{0a}\leftarrow$ \verb"approximateFun"$(W_0,\mu)$;
\STATE $W(k_1) \leftarrow W_{0a}$;$//$ put the merged output to the $k_1$th section of $W$
\STATE $//$ Obtain the transition probability defined in (\ref{eq.W22})
\STATE $W_1\leftarrow$ \verb"calcTran_typeOne"$(W_u, W_b)$;
\STATE $//$ Merge $W_{1}$ with a fixed output alphabet size $\mu$.
\STATE $W_{1a}\leftarrow$ \verb"approximateFun"$(W_1, \mu)$;
\STATE $W(k_2) \leftarrow W_{1a}$;$//$ put the merged output to the $k_2$th section of $W$.
\ENDFOR
\ENDFOR
\STATE $P_e=$\verb"calcErrorProb"$(W)$;$//$ calculate the error probability of all bit channels according to their transition probabilities.
\end{algorithmic}
\end{algorithm}

\begin{algorithm}
\cprotect\caption{Function $(W_u, W_b, k_1, k_2) = $\verb"tran"$(W, i, j)$ $\\$obtain the channel transition probability of the $j$th Z-shape connection at level $i$}
\label{algorithm_transition_prob}
\begin{algorithmic}[1]
\REQUIRE $i$: level $i$ \\
~~~~~$j$:  the Z-shape index \\
~~~~$W$: transition probability of the underlying channel $~~~~~~~~~~W=(W^{(1)}, W^{(2)}, ..., W^{(N)})$;
\ENSURE $W_u$:  upper right channel corresponding to $W$ in (\ref{eq.W21}) \\
$W_b$: the bottom right channel corresponding to $Q$ in (\ref{eq.W22}).\\
$k_1$: index of $W_u$ in the vector $W$ \\
$k_2$: index of $W_b$ in the vector $W$
\STATE $p_{z}=\lceil \frac{j}{2^{i-1}}\rceil$;
\STATE $k_1 = (p_{z}-1)2^{i} + j-(p_z-1)2^{i-1}$;
\STATE $k_2 = (p_z-1)2^{i} + j-(p_z-1)2^{i-1}+2^{i-1}$;
\STATE $W_u = W(k_1) =  W^{(k_1)}$; $//$ the $k_1$th section of $W$
\STATE $W_b = W(k_2) = W^{(k_2)}$; $//$ the $k_2$th section of $W$
\end{algorithmic}
\end{algorithm}

\subsection{Differences and Complexity Analysis Compared with Tal-Vardy's Procedure}
Originally, the $N$ independent underlying channels are identical: $N$ independent copies
of BMS channel $W$. The $N$ independent underlying channels of the general construction in Section
\ref{sec_modified_construction} can be different. The Z-shape (or the one-step transformation)
of the general construction takes the form in Fig.~\ref{fig.basic structure}.
The main difference of the modified Tal-Vary algorithm with the original algorithm lies in the number of calculations of the one-step transformation. In the original Tal-Vardy algorithm, all input channels to Z-shapes of the same group are the same at each level, requiring one calculation of the one-step transformation for each group (Please refer to Section \ref{sub.2.1} for this discussion). Therefore, for the original Tal-Vardy algorithm, the number of calculations of the one-step transformation is $2^{k-1}$ for level $k$.
Suppose all output channels
have size $\mu$. Let us consider the approximate process by which all
channels are stored from level $n$ to level $1$. All $N$ bit channels can be approximated at level $n$.
At level $k$, the memory space to store these channels is thus $2^{k-1}\times 2\times\mu$, leading to the largest memory
space of $2^{n-1}\times2\times\mu = \mu N$.
The total one-step transformation in $n$ levels for the original Tal-Vardy algorithm is therefore: $2^n-1=N-1$, requiring the approximate procedure
to be applied $2(N-1)$ times.
By contrast, in the modified Tal-Vardy algorithm, Z-shapes in the same group in each level can have different input channels, leading to a complete calculation of all one-step transformations in each group. The approximate procedure is
applied to each of the one-step transformations in each level. The total number of one-step transformations is $N\log_2 N$.
These one-step transformations require $N\log N$ approximate procedures and $\mu N\log N$ memory space.
Table \ref{table_complexity} is a summary of the complexity discussion.

\begin{table}[!t]
\renewcommand{\arraystretch}{1.3}
\caption{Complexity Comparison}
\centering
\footnotesize
\label{table_complexity}
\begin{tabular}{c|m{2cm}<{\centering}|m{3cm}<{\centering}}
\Xhline{1pt}
  ~ & \textbf{Largest Memory Space} & \textbf{Number of Approximate Procedures} \\
\hline
 {\textbf{Tal-Vardy}} & $\mu N$ & $2(N-1)$ \\
\hline
 {\textbf{Modified Tal-Vardy}} & $\mu N\log N $  &  $N \log N$\\
\Xhline{1pt}
\end{tabular}
\end{table}

\section{Simulation Results of the General Construction} \label{sec_numerical}
In this section, the construction of polar codes in
BEC channels and the modified Tal-Vardy algorithm in Algorithm \ref{al.ag2} for all other channels
are used to construct polar codes with puncturing and shortening,
echoing our motivation of this paper's work in Section \ref{sub.2.2}.
In the puncturing mode, the receiver has no knowledge of the punctured bits.
{{
The punctured coded bits are not transmitted and the corresponding punctured
channels are modeled as the channel in Lemma \ref{lemma_puncture_channel}.
}}
For BEC channels, this puncturing is equivalent to receiving
an erasure bit at the punctured position.
In AWGN channels with the BPSK modulation, it is
equivalent to a received value of zero at the punctured position.
With shortening,  the shortened bits are known at the decoder side
and are all set to zero in our simulations.
{{
The transmission process is therefore modeled in the following steps:
\begin{itemize}
\item Construction preparation:
\begin{itemize}
    \item According to the puncturing/shortening mode, the punctured/shortened coded bits are obtained.
    \item The corresponding channels are modeled  as either the channel in Lemma \ref{lemma_puncture_channel}
    or the channel in Lemma \ref{lemma_shortened_channel}.
    \item The $N$ underlying channels are obtained: $N-P$ channels are independent and identical copies of the original
    underlying channel $W$, and $P$ channels are either the channel from Lemma \ref{lemma_puncture_channel} or the channel
    from Lemma \ref{lemma_shortened_channel}.
    \item From these $N$ underlying channels, the construction algorithm can be employed
    to re-order the bit channels. For BEC channels, the recursive equations in (\ref{eq.BEC_1}) and (\ref{eq.BEC_2}) can be
    employed to calculate the Bhattacharyya parameters of the bit channels. For AWGN channels, GA \cite{niu_icc13,zhang_it14} or the proposed modified Tal-Vardy procedure can be employed to re-order the bit
    channels. For all other BMS channels, the proposed Tal-Vardy procedure can be employed since GA is no longer applicable.
    \item Denote the information set as $\mathcal{A}$ and the frozen set as $\mathcal{\bar{A}}$ from the previous re-ordering
    of the bit channels.
\end{itemize}
\item The decoding process:
    \begin{itemize}
    \item The $N$ initial LR values can be calculated according to their received symbols and their channel types: the original underlying channel $W$, or the punctured/shortened channel from Lemma \ref{lemma_puncture_channel}/Lemma \ref{lemma_shortened_channel}.
    \item The successive cancellation (SC) \cite{arikan_it09} decoding process carries over from the initial LR values, the information set $\mathcal{A}$, and the frozen set $\mathcal{\bar{A}}$, just as in the original SC decoding process.
    \end{itemize}
\end{itemize}
}}

\begin{figure}
{\par\centering
\resizebox*{3.0in}{!}{\includegraphics{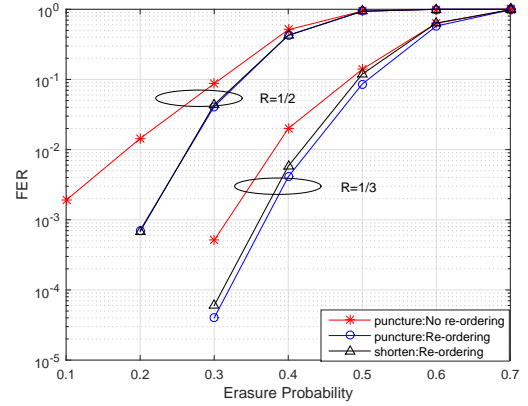}} \par}
\caption{The error probability of polar codes in BEC channels. The original code length is $N=256$. After puncturing and shortening,
the code length is $M=186$ with a code rate $R=1/2$ and $R=1/3$. }
\label{fig.BEC-Error Probability}
\end{figure}

Fig.~\ref{fig.BEC-Error Probability} shows the error probability of polar codes with $N=256$ , $R=1/2$ and $R=1/3$ in the
BEC channels.
The number of punctured (or shortened)  bits is
$P=70$, resulting in a final code length of $M=186$. Define  two vectors: $V_p=(1,2,...,P)$ and $V_s=(N-P+1, N-P+2, ..., N)$.
The punctured and shortened coded bits are bit-reversed versions of $V_p$ and $V_s$, respectively,
as in \cite{niu_icc13,DBLP:journals/corr/NiuD0LZV16}.
In Fig.~\ref{fig.BEC-Error Probability}, the frame-error-rate (FER) performance
is shown (a frame is a code block) where the x-axis is the erasure probability of
the underlying BEC channels. The label with `puncture: No re-ordering' indicates that even with puncturing,
the good bit channels are selected from the  original sorting of bit channels  as though there were no puncturing.
The label with `puncture: Re-ordering' indicates that the bit channels are re-selected from the Bhattacharyya parameters recursively calculated according to
equations (\ref{eq.BEC_1}) and  (\ref{eq.BEC_2}).
The label with `shorten: Re-ordering' corresponds to
the FER performance of reordering bit channels when shortening is performed.
The initial Bhattacharyya parameters at the punctured
positions are set to one. The initial Bhattacharyya parameters at the shortened
positions are set to  zero. It can be observed that re-ordering bit channels with puncturing and shortening
 improves the FER performance of the polar codes.

\begin{figure}
{\par\centering
\resizebox*{3.0in}{!}{\includegraphics{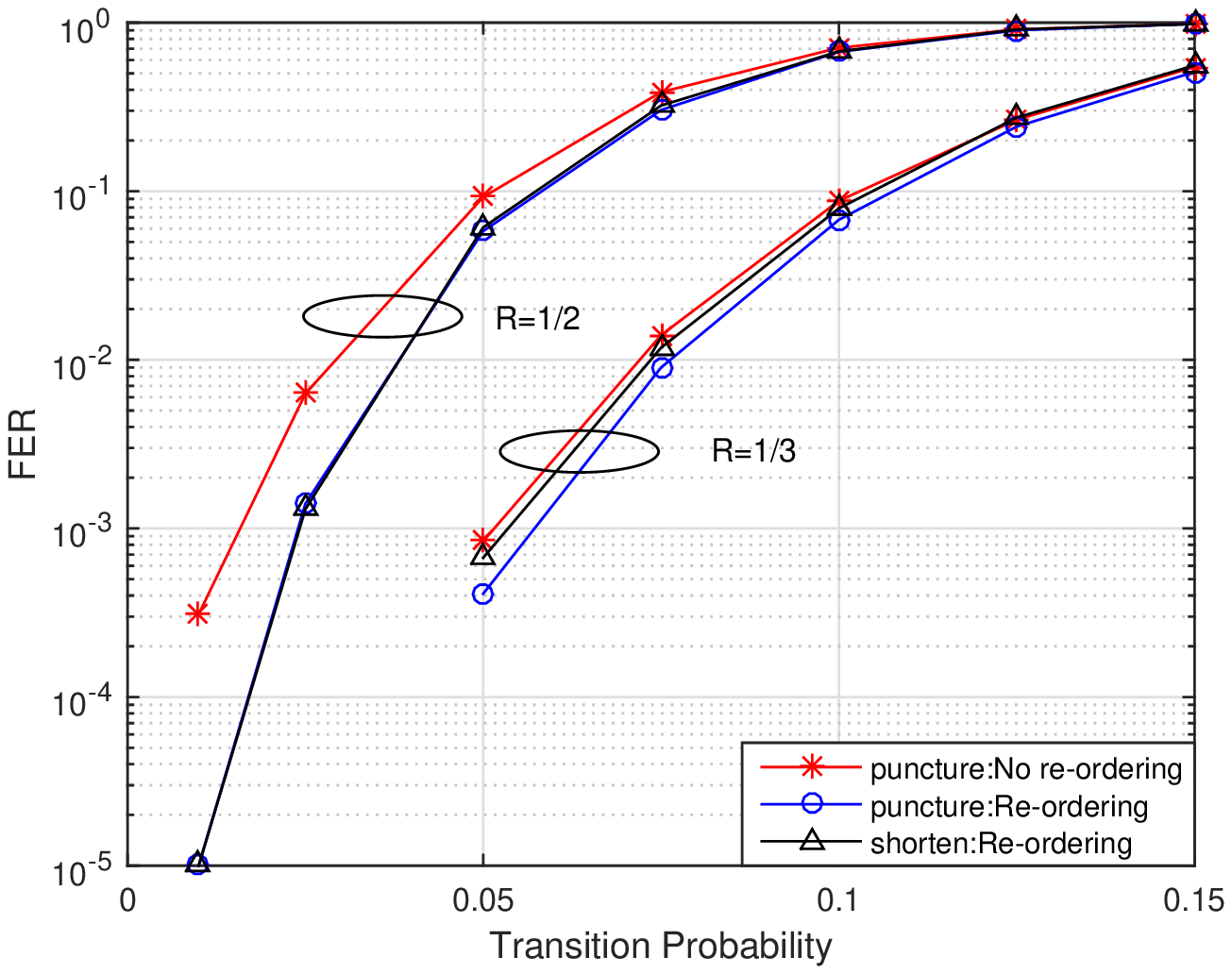}} \par}
\caption{The error probability of polar codes in BSC channels. The original code length is $N=256$. After puncturing and shortening,
the code length is $M=186$ with a code rate $R=1/2$ and $R=1/3$. }
\label{fig.BSC-Error Probability}
\end{figure}

Fig.~\ref{fig.BSC-Error Probability} shows the error performance of puncturing and shortening in binary symmetric channels (BSC).
Puncturing and shortening are conducted in the same fashion as that in the
BEC channels with the same parameters. The x-axis is the transition probability
of the underlying BSC channels.
Originally, the bit channels are sorted according to Tal-Vardy's
degrading merging procedure with $\mu=256$ \cite{vardy_it13} as though there was
no puncturing or shortening. The performance of such a construction is shown in Fig.~\ref{fig.BSC-Error Probability}
by the lines with asterisks (with legend `puncture: No re-ordering').
Applying Algorithm \ref{al.ag2} to re-order the bit channels,
the output alphabet size in each approximate process is still set as $\mu=256$.
At the punctured positions, the transition probabilities are set to $W(y_i|0) = 0.5$ and $W(y_i|1) = 0.5$
(position $i$ corresponds to a punctured position).
At the shortened positions, the transition probabilities are $W(y_i|0) = 1$ and $W(y_i|1) = 0$ since
the shortened bits are set to zero.
As shown in Fig.~\ref{fig.BSC-Error Probability}, if the bit channels are not re-ordered, the FER performance
is degraded compared with the performance with re-ordering.

Note that  polar codes with puncturing or shortening in BSC channels can not be
constructed using Gaussian approximation \cite{dai_ia17,wu_icl14}. However,
the proposed modified Tal-Vardy procedure can be
employed to construct polar codes with independent BMS channels.

\begin{figure}
{\par\centering
\resizebox*{3.0in}{!}{\includegraphics{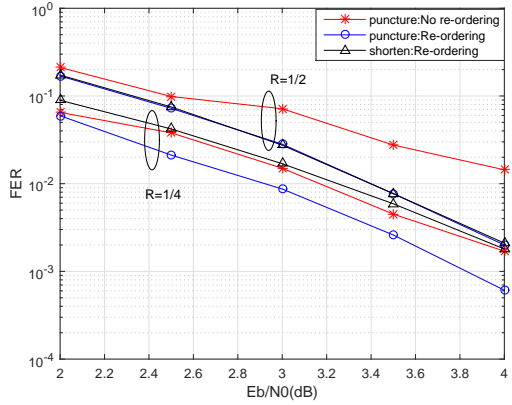}} \par}
\caption{The error  performance of polar codes in AWGN channels. The original code length is $N=256$.
After puncturing and shortening, the code length is $M=186$ with a code rate $R=1/2$ and $R=1/4$.}
\label{fig.AWGN-Error Probability}
\end{figure}

Fig.~\ref{fig.AWGN-Error Probability} shows the error performance of puncturing and shortening in AWGN channels.
The underlying AWGN channel is first converted to a BMS channel
as in \cite{vardy_it13} with an output alphabet size of $2048$.
Then, puncturing and shortening are carried out in the same fashion as the
BEC channel with the same puncturing and shortening parameters. The bit channels are first sorted according to Tal-Vardy's
degrading merging procedure with $\mu=256$ (assuming no puncturing or shortening).
The FER performance of the polar code constructed from this sorting is the lines with asterisks in Fig.~\ref{fig.AWGN-Error Probability} (corresponding to a code rate of $R=1/2$ and $R=1/4$). Applying Algorithm \ref{al.ag2}, the bit channels
are  re-ordered, and the output alphabet size is still $\mu=256$.
At the punctured positions, the channels are treated as receiving a zero (or with a LR of 1);
at the shortened positions, the channels are treated as a known channel with an LR of infinity.
As shown in Fig.~\ref{fig.AWGN-Error Probability},  without re-ordering of the bit
channels, the FER performance is degraded compared with the performance with re-ordering.


\section{Hardware Implementation}\label{sec:vlsi}
In this section, a general folded polar encoder architecture is studied. Based on that, a pruned architecture is proposed which could reduce the latency and improve the throughput significantly.
\subsection{General Folded Polar Encoder}
Folding is a transformation technique to reduce the hardware area by multiplexing the processing time. When polar codes are applied in ultra-reliable scenarios, the code length is long and the hardware area is large. By exploiting the similarity between polar encoding and the FFT, the folded polar encoders are proposed in \cite{yoo2015partially}, \cite{zhang2015pipelined}.

\begin{figure}[htbp]
  \centering
  \includegraphics[width=0.7\linewidth]{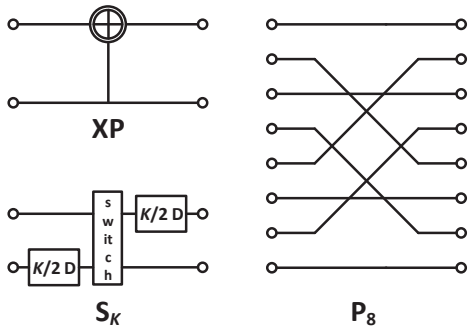}\\
  \caption{Basic module of the folded polar encoder.}\label{fig:sym}
\end{figure}

Illuminated by \cite{zhong2018auto}, a general description of folded polar encoder is summarized which requires less registers than \cite{zhong2018auto}. The folded architecture could be represented by a general equation. The folded polar encoder is composed of three basic modules which are shown in Fig. \ref{fig:sym}. The XOR-or-PASS is the arithmetic unit and is abbreviated as \textbf{XP}. The module \textbf{S}$_K$ is the commutator with $K/2$ delays which switches the signal flow. The \textbf{P}$_K$ is the permutation module with $K$ inputs and $K$ outputs. Fig. \ref{fig:sym} shows the case of \textbf{P}$_8$.

\begin{figure}[htbp]
  \centering
  \includegraphics[width=0.95\linewidth]{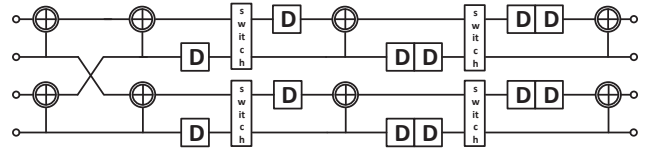}\\
  \caption{The folded polar encoder architecture for $N=16$ and $L=4$.}\label{fig:p16-4}
\end{figure}

Suppose that the parallelism degree is $L$ and the overall architecture could be represented as below,
\begin{equation}
\mathrm{Arch}\! = \!\prod\limits_{i=0}^{\log_2L-1}\left(\textbf{XP}^{\otimes L/2}\cdot\!\!\textbf{P}_{L/2^i}^{\otimes 2^i}\right)\cdot\prod\limits_{i=1}^{\log_2(N/L)}\left(\textbf{S}_{2^i}^{\otimes L/2}\right),
\end{equation}
where $\ast^{\otimes m}$ means the module has $m$ copies at each stage. For example, when $N=16$ and $L=4$, the architecture is shown in Fig. \ref{fig:p16-4}.

Based on the equation, the hardware code of the folded polar encoder could be generated automatically. In this paper, this architecture is abbreviated as ``auto encoder''.
\subsection{Pruned Folded Polar Encoder}
\begin{figure}[htbp]
  \centering
  \includegraphics[width=0.9\linewidth]{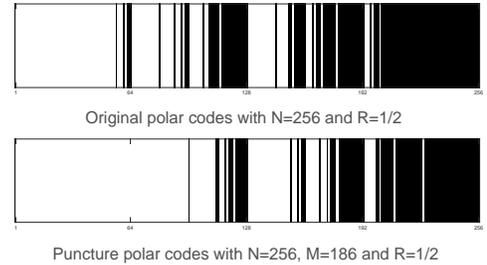}\\
  \caption{The bit distribution of original polar codes with $N=256$ and $R=1/2$, and punctured polar codes with $N=256$, $M=186$ and $R=1/2$. The white part contains frozen bits (bits $0$s) and the black lines are information bits (bits $1$s) in the two examples. }\label{fig:dis}
\end{figure}

Based on the partial orders \cite{mori_icl09,schurch_isit16,li_iwcmc17} of polar codes, it can be concluded that the bit channels of polar codes with large indices tend to be good bit channels; and those with small indices tend to be frozen bit channels. Of course there are regions where frozen and information bits are tangled together. For example, in Fig.~\ref{fig:dis}, the beginning part of the source bits contains frozen bits (bits $0$s) and the last part is the information bits (bits $1$s). With the puncturing such as \cite{niu_icc13}, the first $P$ bit channels are punctured, meaning that there are consecutive $Q$ frozen bits $0$ at the beginning of the source bits. The bottom figure of Fig.~\ref{fig:dis} shows an example of the distribution of the frozen and information bits. It is intuitive that the beginning part of number of $0$s (frozen bits) increases with puncturing.

Since the XOR value of two `$0$'s is still `$0$', the process for the beginning bits could be avoided. In $L$-parallelism folded polar encoder, the source bits enters the circuit in blocks of length $L$. Define the latency as the first data-in to the first data-out. The latency of the encoder is $\frac{N}{L}$. According to the bit distribution, denote $C$ as the number of $0$s at the beginning. For the punctured polar code in Fig. \ref{fig:dis}, $C =95$. The data-in for these bits could be pruned. Herein, the latency could be reduced to $\left\lceil\frac{N-C}{L}\right\rceil$. Correspondingly, the last $\left\lceil\frac{C}{L}\right\rceil$ data-out cycles should be combined to one cycle.
\begin{figure*}[htbp]
  \centering
  \includegraphics[width=0.8\linewidth]{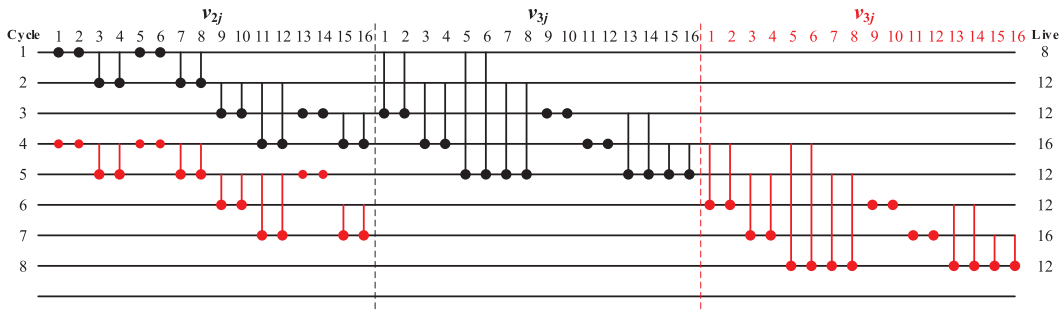}\\
  \caption{Life analysis of the pruned folded encoder.}\label{fig:l1}
\end{figure*}

Take Fig. \ref{fig:p16-4} as an example, label the data-in and data-out as $\mathbf{u_{1}}$ to $\mathbf{u_{4}}$ and $\mathbf{x_{1}}$ to $\mathbf{x_{4}}$. The clock cycles of the original folded polar encoder and pruned folded polar encoder are illustrated in Table \ref{tab:cc}. It can be seen that the latency of the original encoder is $4$ when pipelined, while the latency of the pruned encoder is reduced to $3$ when pipelined.
\begin{table}[htbp]
  \centering
  \tabcolsep 1.6mm
  \footnotesize
  \renewcommand{\arraystretch}{1.3}
  \caption{Clock cycles comparison between the original folded encoder and the pruned folded encoder.}
    \begin{tabular}{c|c||cccccccc}
    \Xhline{1pt}
                                         &                                   & \multicolumn{8}{c}{\textbf{Clock Cycle}}                                                                                                                                                                                                                                                                                                            \\ \cline{3-10}
    \multirow{-2}{*}{\textbf{Encoder}}   & \multirow{-2}{*}{\textbf{Signal}} & 1                                      & 2                                      & 3                                      & 4                                      & 5                                                     & 6                                      & 7                                      & 8                                      \\ \hline
                                         & \textbf{in}                       & \cellcolor[HTML]{DEDBDB}$\mathbf{u}_1$ & \cellcolor[HTML]{DEDBDB}$\mathbf{u}_2$ & \cellcolor[HTML]{DEDBDB}$\mathbf{u}_3$ & \cellcolor[HTML]{DEDBDB}$\mathbf{u}_4$ & $\mathbf{u}_1$                                        & $\mathbf{u}_2$                         & $\mathbf{u}_3$                         & $\mathbf{u}_4$                         \\ 
    \multirow{-2}{*}{\textbf{Original}}  & \textbf{out}                      &                                        &                                        &                                        & \cellcolor[HTML]{DEDBDB}$\mathbf{x}_1$ & \cellcolor[HTML]{DEDBDB}$\mathbf{x}_2$                & \cellcolor[HTML]{DEDBDB}$\mathbf{x}_3$ & \cellcolor[HTML]{DEDBDB}$\mathbf{x}_4$ & $\mathbf{x}_1$                         \\ \hline
                                         & \textbf{in}                       & \cellcolor[HTML]{DEDBDB}$\mathbf{u}_2$ & \cellcolor[HTML]{DEDBDB}$\mathbf{u}_3$ & \cellcolor[HTML]{DEDBDB}$\mathbf{u}_4$ & $\mathbf{u}_2$                         & $\mathbf{u}_3$                                        & $\mathbf{u}_4$                         & \cellcolor[HTML]{DEDBDB}$\mathbf{u}_2$ & \cellcolor[HTML]{DEDBDB}$\mathbf{u}_3$ \\ 
    \multirow{-2}{*}{\textbf{Pruned}} & \textbf{out}                      &                                        &                                        & \cellcolor[HTML]{DEDBDB}$\mathbf{x}_1$ & \cellcolor[HTML]{DEDBDB}$\mathbf{x}_2$ & \cellcolor[HTML]{DEDBDB}$\mathbf{x}_3$,$\mathbf{x}_4$ & $\mathbf{x}_1$                         & $\mathbf{x}_2$                         & $\mathbf{x}_3$,$\mathbf{x}_4$          \\ \Xhline{1pt}
    \end{tabular}
  \label{tab:cc}%
\end{table}%

The corresponding life analysis chart is drawn in Fig. \ref{fig:l1}. The black line represents the first frame and the red is the second frame. The variable $v_{i,j}$ denotes the temporary value of the $j$th row of the encoding structure in the $i$th stage. In the pruned encoder, $\{v_{2,1},v_{2,2},v_{2,5},v_{2,6}\}$ are known at the beginning and the input starts from $\mathbf{u}_2$. Moreover, $v_{3,5-8}$ and $v_{3,13-16}$ are encoded and outputted together.

\begin{figure}[htbp]
  \centering
  \includegraphics[width=1\linewidth]{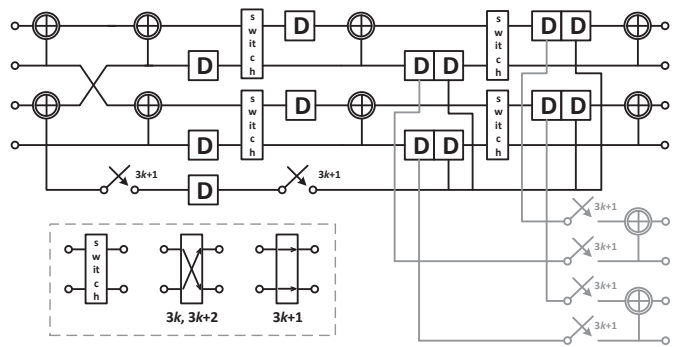}\\
  \caption{The pruned hardware architecture for $N=16$ and $L=4$.}\label{fig:punc}
\end{figure}

\begin{table*}[htbp]
  \centering
  \tabcolsep 2.4mm
  \footnotesize
  \renewcommand{\arraystretch}{1.4}
  \caption{XOR gates, registers, latency and throughput comparisons between original encoder and pruned encoder.}
    \begin{tabular}{l|c||c|c|c|c}
    \Xhline{1pt}
      \multicolumn{2}{c||}{}  & \textbf{XOR gates}         & \textbf{Registers} & \textbf{Latency} [cycle]                & \textbf{Throughput} [bit/cycle]           \\ \hline
    \multirow{2}{*}{\textbf{Original Encoder}} & \textbf{Y. Hoo} \cite{yoo2015partially} & $L/2 \log_2N$     & $N-L$     & $N/L$                          & $L$                              \\ \cline{2-6}
                                      & \textbf{Z. Zhong} \cite{zhong2018auto} & $L/2 \log_2N$     & $3N/2-L$  & $3N/2L-1$                          & $\frac{2LN}{3N-2L}$                              \\ \hline
    \multicolumn{2}{l||}{\textbf{Pruned Encoder}}    & $L/2 (\log_2N+C)$ & $N-L+2$   & $\lceil \frac{N-C}{L}\rceil$ & $N/\lceil \frac{N-C}{L}\rceil$ \\ \Xhline{1pt}
    \end{tabular}
  \label{tab:comp}%
\end{table*}%
\begin{figure}[htbp]
  \centering
  \includegraphics[width=1\linewidth]{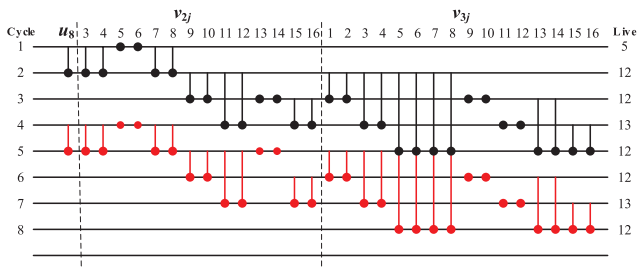}\\
  \caption{Optimized life analysis of the pruned folded encoder.}\label{fig:l2}
\end{figure}
The pruned encoder costs $4$ more registers which store $v_{3,1},v_{3,2},v_{3,5},v_{3,6}$ than the original at the $4$th cycle. According to the \textbf{XP} unit, if the last bit of data-in is an information bit and the rest are frozen, all encoding bits of this set are equalling to the last bit. Therefore, the extra four registers could combine to one register since the four values are all the same as $u_8$. Furthermore, the encoded results of $\mathbf{u}_1$ are directly set to $0$ and stored in the registers. Based on this analysis, Fig. \ref{fig:l1} could be further optimized to Fig. \ref{fig:l2}. It can be concluded that the number of registers increases only by $1$ but the latency is reduced by $25\%$ for polar codes with $N=16$ and $L=4$. The corresponding architecture is shown in Fig. \ref{fig:punc}.

To sum up, the comparison between the original encoder and the pruned encoder is listed in Table \ref{tab:comp}, including latency, XOR gates, delay elements and throughput. Generally, due to the padding of clock cycles, the number of extra registers should be $\lfloor\frac{C}{L}\rfloor\times L$. Suppose the first data-in has only one information bit $u_{C+1}$, the number of registers could be further reduced to $2$ in most cases. One stores $0$ which is the immediate value of the previous bits and the other stores $u_{C+1}$ for the next cycle.

\subsection{Implementation Results}
The auto encoder and pruned encoder were implemented on the Xilinx Virtex-$7$ XC$7$VX$690$T FPGA platform. The results of $(256,186)$ and $(1024,744)$ polar codes with $R=1/2$ and $L=32$ are given when $E_b/N_0=2$ dB, $C=95$ and $C=342$, respectively. Furthermore, the results are compared with the state-of-the-art. G. Sarkis \cite{sarkis2015flexible} uses the architecture of \cite{yoo2015partially}. Herein, its results are normalized for fair comparison.
\begin{table*}[htbp]
  \centering
  \footnotesize
  \tabcolsep 3.5mm
  \renewcommand{\arraystretch}{1.3}
  \caption{Implementation results comparison with state-of-the-art works.}
    \begin{tabular}{l|ll|llll}
    \Xhline{1.0pt}
     $(N,C)$& \multicolumn{2}{c|}{$(256,186)$} & \multicolumn{4}{c}{$(1024,744)$} \\
    \rowcolor{tabgray} \textbf{Encoder} & \textbf{Auto Encoder} & \textbf{Pruned Encoder} & \textbf{G. Sarkis} \cite{sarkis2015flexible} & \textbf{Z. Zhong} \cite{zhong2018auto} &  \textbf{Auto Encoder}   & \textbf{Pruned Encoder}\\ \hline
     \textbf{LUT}   & $201$   & $227$   &$275$ & $467$   & $359$   & $415$ \\
    \rowcolor{tabgray}\textbf{FF}    & $134$   & $192$ & $789$ & $312$   & $207$   & $324$ \\
    Total & $335$ & $419$ & $1064$ & $779$ & $566$ & $739$ \\ \hline
    \rowcolor{tabgray} \textbf{Latency} [cycle] & $8$   & $6$    & $32$ & $47$      & $32$      & $22$ \\
    \textbf{Max. Freq.} [MHz] & $415.8$ & $390.75$ &$401.23$& $407.05$ & $392.17$ & $359.18$ \\
    \rowcolor{tabgray} \textbf{Throughput} [Gbps] & $13.31$   & $16.67$ &   $12.84$  & $8.87$      & $12.55$      &  $16.07$\\
    \Xhline{1.0pt}
    \end{tabular}%
  \label{tab:addlabel}%
\end{table*}%

\begin{figure}
  \centering
  \includegraphics[width=1.0\linewidth]{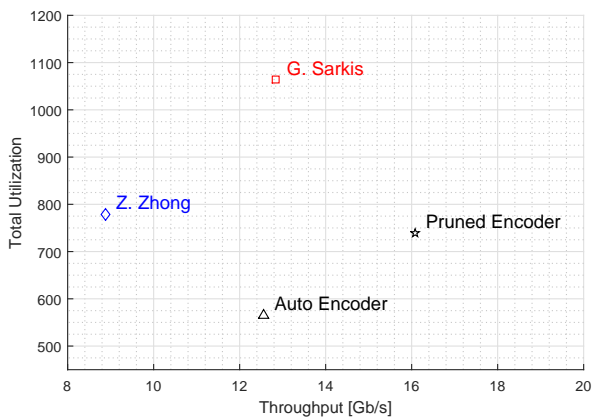}\\
  \caption{Implementation comparison of state-of-the-art partially folded polar encoders.}\label{fig:comp}
\end{figure}

The pruned encoder is implemented based on the auto encoder, which improves the throughput by $25\%$ with additional $25\%$ area cost for $(256,186)$ codes. The additional area is due to the control module. For $(1024,744)$, the implementation comparison chart is plotted in Fig. \ref{fig:comp}. The point which is near the bottom right corner shows the best performance. It can be concluded that pruned encoder outperforms other two works in both utilization and throughput. Compared with the auto encoder, the latency of the pruned encoder is reduced by $31\%$ and the throughput is improved by $28\%$.

\section{Conclusion}
This paper focuses on the construction of  polar codes in general cases; i.e., the underlying
channels are independent BMS channels.
In terms of software, proofs are presented to show that the symmetric property and the degradation relationship are still preserved.
From these theoretical aspects, a general construction of polar codes based on Tal-Vardy's algorithm is proposed.
The general construction can be applied to all types of independent BMS channels.
In terms of hardware, the property of polar codes could optimize the hardware encoder architecture. For the proposed pruned folded encoder, the latency is reduced by $31\%$ and throughput is improved by $28\%$ when the length of code is $1024$.
\appendix[Proof of Lemma \ref{lemma_puncture_channel}]
The input alphabet is $\mathcal{X} = \{0,1\}$, and the output is $\mathcal{Y}=\{y\}$.
From Lemma \ref{lemma_puncture_channel}, the transition probability is $H(y|0)=H({y}|1)=\frac{1}{2}$.
From (\ref{eq_iw}), the symmetric  capacity of this channel $H$ is
\begin{align*}
I(H)&=\sum\limits_{x\in \mathcal{X}}\frac{1}{2}H(y|x)\log{\frac{H(y|x)}{\frac{1}{2}H(y|0)+\frac{1}{2}H(y|1)}}\\
&=\frac{1}{2}H(y|0)\log{\frac{H(y|0)}{\frac{1}{2}H(y|0)+\frac{1}{2}H(y|1)}} \\
&~~~+\frac{1}{2}H(y|1)\log{\frac{H(y|1)}{\frac{1}{2}H(y|0)+\frac{1}{2}H(y|1)}}\\
&=0.
\end{align*}
Therefore, the symmetric capacity of the channel $H$ is $I(H)=0$.

\bibliography{IEEEabrv,ref_polar}
\bibliographystyle{IEEEtran}


\end{document}